\documentclass{vhepu14}

\usepackage{amsmath}
\usepackage{amssymb}

\def\be{\begin{equation}}
\def\ee{\end{equation}}
\def\bea{\begin{eqnarray}}
\def\eea{\end{eqnarray}}

\newcommand{\Photo}{}

\begin{document}
\vspace*{4cm}
\title{EXTRAGALACTIC PROPAGATION OF\\ ULTRA-HIGH ENERGY COSMIC RAYS}

\author{ DANIEL KUEMPEL }

\address{RWTH Aachen University, III. Physikalisches Institut A, \\
Otto-Blumenthal-Stra\ss e, 52056 Aachen, Germany}

\maketitle\abstracts{
More than 100 years after the discovery of cosmic rays and various experimental efforts, the origin of ultra-high energy cosmic rays ($E > 10^{17}$~eV) remains unclear. The understanding of production and propagation effects of these highest energetic particles in the universe is one of the most intense research fields of high-energy astrophysics. With the advent of advanced simulation engines developed during the last couple of years, and the increase of experimental data, we are now in a unique position to model source and propagation parameters in an unprecedented precision and compare it to measured data from large scale observatories. In this paper we revisit the most important propagation effects of cosmic rays through photon backgrounds and magnetic fields and introduce recent developments of propagation codes. Finally, by comparing the results to experimental data,  possible implications on astrophysical parameters are given.}

\section{Introduction}
The observation of cosmic rays with ultra-high energies poses interesting questions. Even more than 50 years after the detection of particles of 100~EeV~\cite{Linsley63} (1~EeV $= 10^{18}$~eV) many issues are still unanswered. What mechanism in the universe can accelerate particles to such high energies? What is their origin and what kind of particles are they? What can they tell us about fundamental and particle physics? Is there a maximal energy they can reach? To tackle these problems large-scale observatories have been built at various locations enabling the observation of different parts of the sky. Today, the most prominent sites are the Pierre Auger Observatory~\cite{PAO2004,PAO2010} in the southern hemisphere located in the Argentinean Pampa Amarilla and the Telescope Array~\cite{TA2012} (TA) in the northern hemisphere in Millard County, Utah, USA. The measurement of the particle flux, elemental composition, arrival directions and temporal variations are of central importance to get a clue of an answer. However, to interpret the observations a detailed knowledge of particle propagation effects is essential. In fact, the propagation of ultra-high energy cosmic rays (UHECR) from the source to the observer modifies the original source spectra and chemical composition due to interactions with low energy photons and matter. Propagation also influences the sky distribution of arriving charged cosmic rays due to deflections in cosmic magnetic fields. The open question of the chemical composition of highest-energy cosmic rays is in fact linked to the question on the size of deflection in cosmic magnetic fields. Any consistent interpretation of the nature and origin of ultra-high energy cosmic rays thus has to include propagation in a three-dimensional environment. With the advent of advanced simulation engines, described in more detail in Sec.\ \ref{sec:simulation}, we are now in a unique position to model source and propagation parameters in an unprecedented precision using computer clusters.\\

The paper is organized as follows: in Sec.\ \ref{sec:Prop} we will revisit the most important interaction processes for ultra-high energy cosmic rays and their secondaries. Deflections in galactic and extragalactic magnetic fields are discussed in Sec.\ \ref{sec:mag}. The current status of public propagation codes is given in Sec.\ \ref{sec:simulation}. Finally, some prospects are given in Sec.\ \ref{sec:MPChall} on comparing results of simulations with measurements from large-scale observatories before concluding in Sec.\ \ref{sec:con}.

\section{Propagation effects en route to Earth}
\label{sec:Prop}
\begin{figure}
\centerline{\includegraphics[width=0.85\linewidth, draft=false]{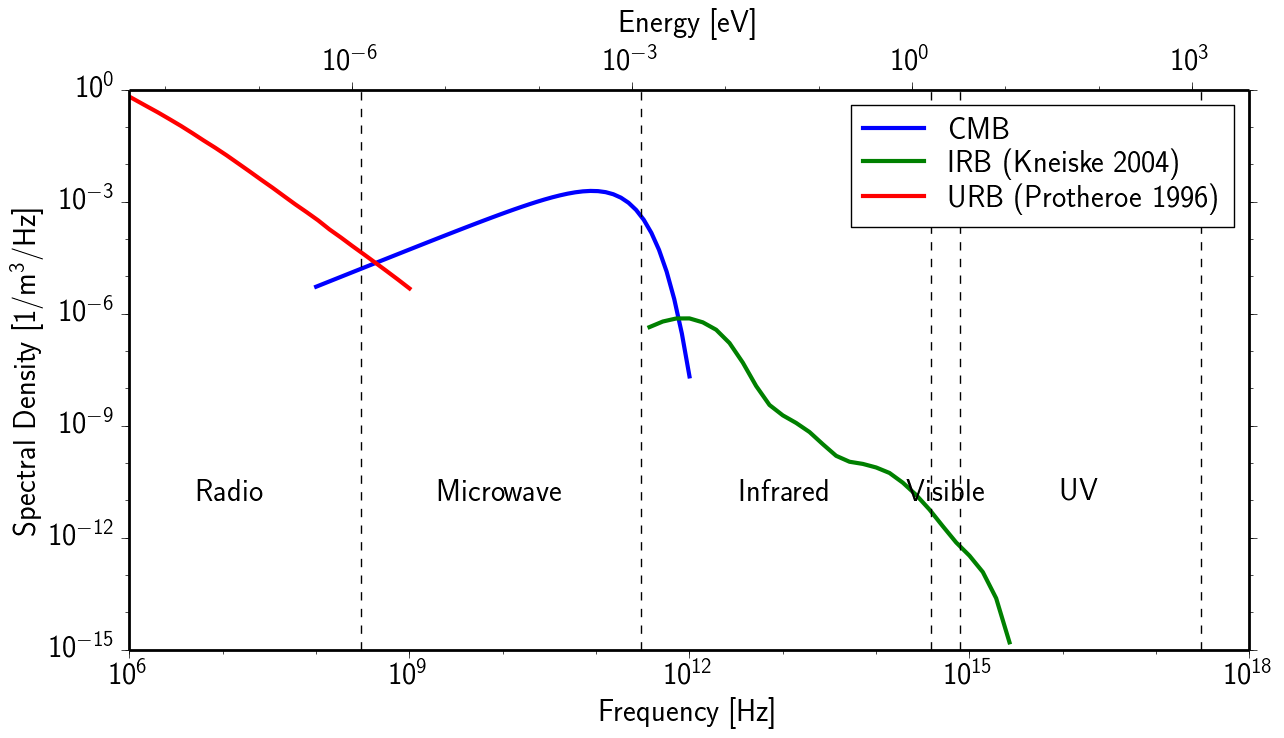}}
\caption[]{Spectral density of the CMB (blue line), infrared background (IRB, green line)~\cite{Kneiske:2003tx}, and universal radio background (URB, red line)~\cite{Protheroe:1996si} as a function of energy and frequency.}
\label{fig:density}
\end{figure}

In the following we shall review the most important aspects of particle propagation through the universe. Except in very close vicinity to the source, only background photons and magnetic fields are relevant to estimate the interactions. In the intergalactic medium the most important photon background is the cosmic microwave background radiation (CMB) with a typical energy of about $10^{-3}$~eV. In addition, cosmic rays can interact with optical and infrared backgrounds as well as with radio waves. The spectral density in the interstellar medium is shown in Fig.\ \ref{fig:density}.
As a consequence of the high energies of cosmic rays with Lorentz factors $\Gamma = E/m$, background photons are seen highly blue-shifted in the nucleus rest frame with the relevant energy $\epsilon' = \epsilon \Gamma (1-\cos\vartheta)$, with the photon energy $\epsilon$ in the laboratory frame and the collision angle $\vartheta$. The interaction length $\lambda$ of a cosmic ray through an isotropic background can be calculated as 
\begin{equation}
\lambda^{-1}(E) = \int_0^\infty n(\epsilon) \sigma_{\rm avg}(\epsilon)~{\rm d}\epsilon~~,
\end{equation}
where $n(\epsilon)$ is the spectral number density of the background particles (cf.\ Fig.\ \ref{fig:density}) and $\sigma_{\rm avg}(\epsilon)$ the cross section for the relevant process averaged over all collision angles $\vartheta$. 

\subsection{Photo-pion production}
\label{sec:pion}
The pion production for a head-on collision of a nucleon $N$ with a background photon $\gamma$ can be described as $N + \gamma \longrightarrow N + \pi$, with a threshold energy of 
\begin{equation}
E_{\rm thres}^{N,\pi} = \frac{m_\pi(m_N + m_\pi /2)}{2\epsilon} \approx 6.8 \cdot 10^{19}~\left( \frac{\epsilon}{10^{-3}~{\rm eV}} \right)^{-1} {\rm eV}~~,
\end{equation}
where $m_{\pi}$ and $m_N$ are the masses of the pion and the nucleon and $\epsilon \sim 10^{-3}$~eV represents a typical energy of a CMB photon. Due to the high inelasticity~\footnote{the inelasticity is typically $\eta = 0.2$ close to the threshold and $\eta = 0.5$ far above the threshold.} $\eta$ of the process and the dense CMB photons it was already realized in the 1960s by Greisen~\cite{Greisen1966} and Zatsepin \& Kuz'min~\cite{Zatsepin1966}, that the universe is opaque for ultra-high energy particles, leading to the so-called GZK flux suppression. A prominent example of photo-pion production by protons is given by 
\begin{equation}
p + \gamma \rightarrow \Delta^+ \rightarrow \left\{   \begin{array}{l l}    n + \pi^+ & \quad \text{with branching ratio 1/3}\\  p + \pi^0 & \quad \text{with branching ratio 2/3} \end{array} \right.,
\end{equation}
where a proton interacts electromagnetically with a photon and excites the proton to the $\Delta^+$ resonance before decaying via strong interactions. In the channel that conserves the charge of the original nucleon mostly neutral pions are produced which decay into secondary gamma rays $\pi^0 \rightarrow \gamma + \gamma$, whereas charge exchange reactions produce mostly charged pions which eventually decay into electrons, positrons and neutrinos. In fact, these are the main production channels for ultra-high energy secondary photons and neutrinos by hadronic cosmic rays, cf.\ Sec.\ \ref{sec:secondaries}. \\
Pion production by nuclei can be described in good approximation by the superposition model. Here nuclei are treated as a superposition of $Z$ free protons and $A-Z$ free neutrons~\footnote{the binding energy is neglected.}. Note that the energy carried away by a pion is only $\eta / A$ of the energy of the primary nucleus with an increased threshold of $E_{\rm thres}^{N,\pi} \cdot A$.

\subsection{Pair production}
Another important interaction process is pair production by a nucleus $X$ with mass number $A$ and atomic number $Z$ on a photon $^A_ZX + \gamma \longrightarrow~ ^A_ZX + e^+ + e^-$. This reaction has a threshold energy of 
\begin{equation}
E_{\rm thres}^{e^\pm} = \frac{m_e(m_X + m_e)}{\epsilon} \approx 4.8 \cdot 10^{17}~A~\left( \frac{\epsilon}{10^{-3}~{\rm eV}} \right)^{-1} {\rm eV}~~,
\end{equation}
and a relatively small inelasticity of about $\eta \sim 10^{-3}$. Therefore, pair production is typically treated as a continuous energy-loss process, but is especially important when calculating secondary photons below PeV energies. 

\subsection{Photodisintegration of nuclei}
In a photodisintegration process a photon is absorbed by an atomic nucleus leading to an excited nuclear state before splitting into two or more parts. Depending on the photon energy $\epsilon'$ in the rest frame of the nuclei different processes are dominant. At low photon energies up to 30~MeV, the giant dipole resonance with the emission of one or two nucleons is the most important contribution. At higher energies, between 30~MeV and 150~MeV, the quasi-deuteron process dominates with predominantly multi-nucleon emission. The effective loss rate can be described as 

\begin{equation}
\frac{1}{E} \left. \frac{{\rm d}E}{{\rm d}t}\right|_{\rm eff} = \frac{1}{A}\frac{{\rm d}A}{{\rm d}t} = \sum_i \frac{i}{A} R_{A,i}(E)~
\end{equation}
where $R_{A,i}$ is the rate for emission of $i$ nucleons from a nucleus of mass $A$.

\subsection{Other energy-loss processes}
An important loss process which dominates near or below the pair production threshold are redshift losses due to the expansion of the universe. This adiabatic fractional energy loss can be described as
\begin{equation}
-\frac{1}{E} \left( \frac{{\rm d}E}{{\rm d}t} \right)_{\rm adiabatic} = H_0~~,
\end{equation}
where $H_0$ is the Hubble constant. \\
Another relevant propagation effect is nuclear decay of unstable particles produced e.g.\ by photodisintegration or photo-pion processes. A nuclear decay can change the energy of the particle as well as the nucleus type.\\
\begin{figure}
\centerline{\includegraphics[width=1.05\linewidth, draft=false]{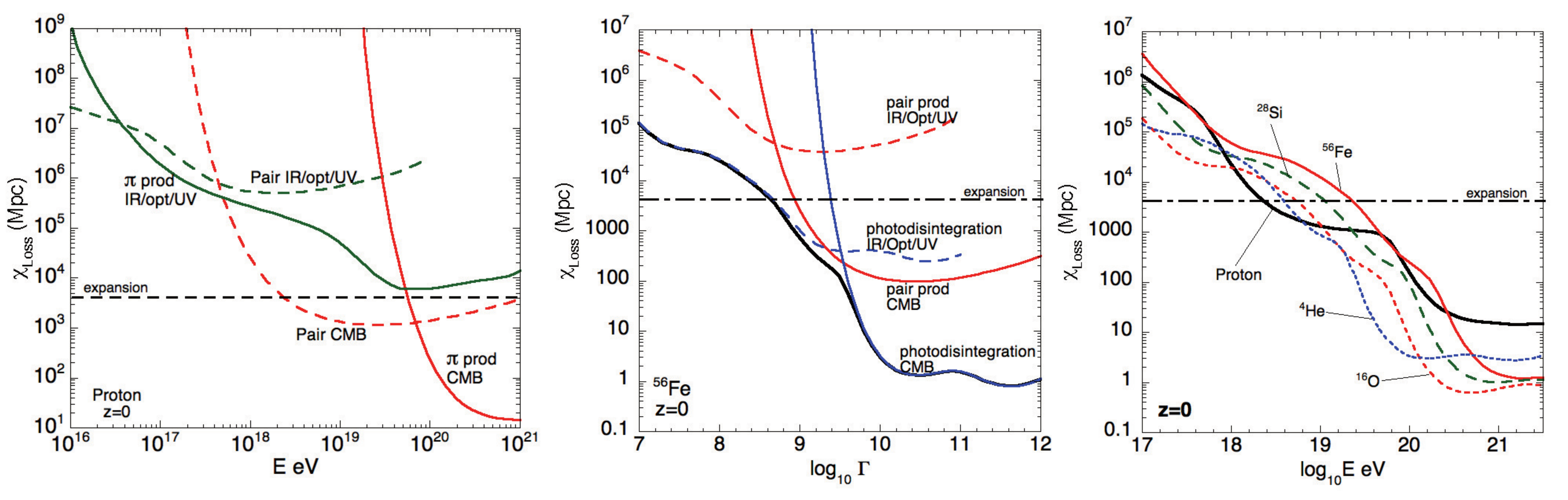}}
\caption[]{Left: Energy-loss length $\chi_{\rm Loss} = \left( \frac{1}{E}\frac{{\rm d}E}{{\rm d}x} \right)^{-1}$ of primary protons as a function of energy. Different energy-loss processes on various photon backgrounds (CMB, infrared, optical and ultra-violet) are indicated. Center: Energy loss length for iron as a function of the Lorentz factor $\Gamma$. Different contributions of pair production and photodisintegration on various backgrounds are indicated. Right: Energy loss length for different nuclei as a function of energy. In all three figures, the effect of adiabatic expansion of the universe is indicated by a horizontal dashed line. (From~\cite{Allard:2011aa})}
\label{fig:loss}
\end{figure}
A graphical illustration of various processes of energy loss for protons as well as for nuclei is shown in Fig.\ \ref{fig:loss}. For protons, the energy loss is dominated below a few EeV by the expansion of the universe. At intermediate energies, pair production on the CMB is most relevant while at energies above $\sim 70$~EeV pion production becomes dominant. For iron nuclei photodisintegration represents the most important loss mechanism at high energies.

\subsection{Secondary photons}
\label{sec:secondaries}
As already discussed in Sec.\ \ref{sec:pion} photo-pion production by protons is the main production channel for ultra-high energy secondary photons. Since photons have no charge, they are not deflected by magnetic fields. However, the existing cosmic photon background creates additional interactions. The dominant process is the attenuation of the ultra-high energy photons due to pair production on background photons, $\gamma_{\rm UHE} + \gamma_{\rm b} \rightarrow e^+ + e^-$. The produced $e^\pm$ can again interact with background photons via inverse Compton scattering resulting in an electromagnetic cascade that ends at GeV-TeV energies where the universe becomes increasingly transparent for photons. Typical energy-loss lengths are $7-15$~Mpc at 10~EeV and $5-30$~Mpc at 100~EeV~\cite{Risse:2007sd}.

\section{Magnetic fields}
\label{sec:mag}
During the propagation charged cosmic rays are deflected by extragalactic and galactic magnetic fields. Considering a particle with charge $Z$ and energy $E$ in [PeV], the Larmor radius $r_L$ in [pc] can be estimated as
\begin{equation}
\left( \frac{r_L}{{\rm pc}} \right) = 1.1 \left(\frac{E}{{\rm PeV}}\right) \left( \frac{{\rm \mu G}}{B} \right) \frac{1}{Z}
\end{equation} 
with the magnetic field $B$ in [$\mu$G]. \\
The parameter space for magnetic fields in the universe is large, since field strengths and especially field orientations are not well constrained. Especially for \textit{extragalactic} magnetic fields predictions vary a lot. Their origin is not well understood~\cite{Kulsrud:2007an} and theories vary from the creation in the primordial universe~\cite{Widrow:2002ud} to magnetic pollution from astrophysical sources (e.g.\ \cite{Scannapieco:2005hk}) such as jets from radio galaxies. Typical strengths are expected to be $\sim 1-40$~$\mu$G in the core of clusters of galaxies~\cite{Olinto2011} and $10^{-16} - 10^{-6}$~G in filaments. The simulation and prediction of large-scale magnetic fields is a dedicated task. Assuming that the fields are induced in galaxies, one would expect stronger fields in high-peaked density regions and a nearly suppressed field in voids. To model more realistic inhomogeneous configurations various groups have developed large-scale structure simulations including magnetic fields, e.g.\ \cite{Sigl:2004yk,Dolag:2004kp,Das:2008vb}. However, these simulations lead to discrepant results due to the variety of assumptions that have to be made. To constrain the strength of extragalactic magnetic fields further observations are needed, e.g.\ via the arrival directions of charged cosmic rays at ultra-high energies, through the observation of extended gamma-ray emission around point sources in connection with the time delay in gamma-ray flares~\cite{Neronov:2009gh}, or through Faraday rotation measurements, e.g.\ with the future Square Kilometer Array~\cite{Beck2007}. 

\begin{figure}
\centerline{\includegraphics[width=1.05\linewidth, draft=false]{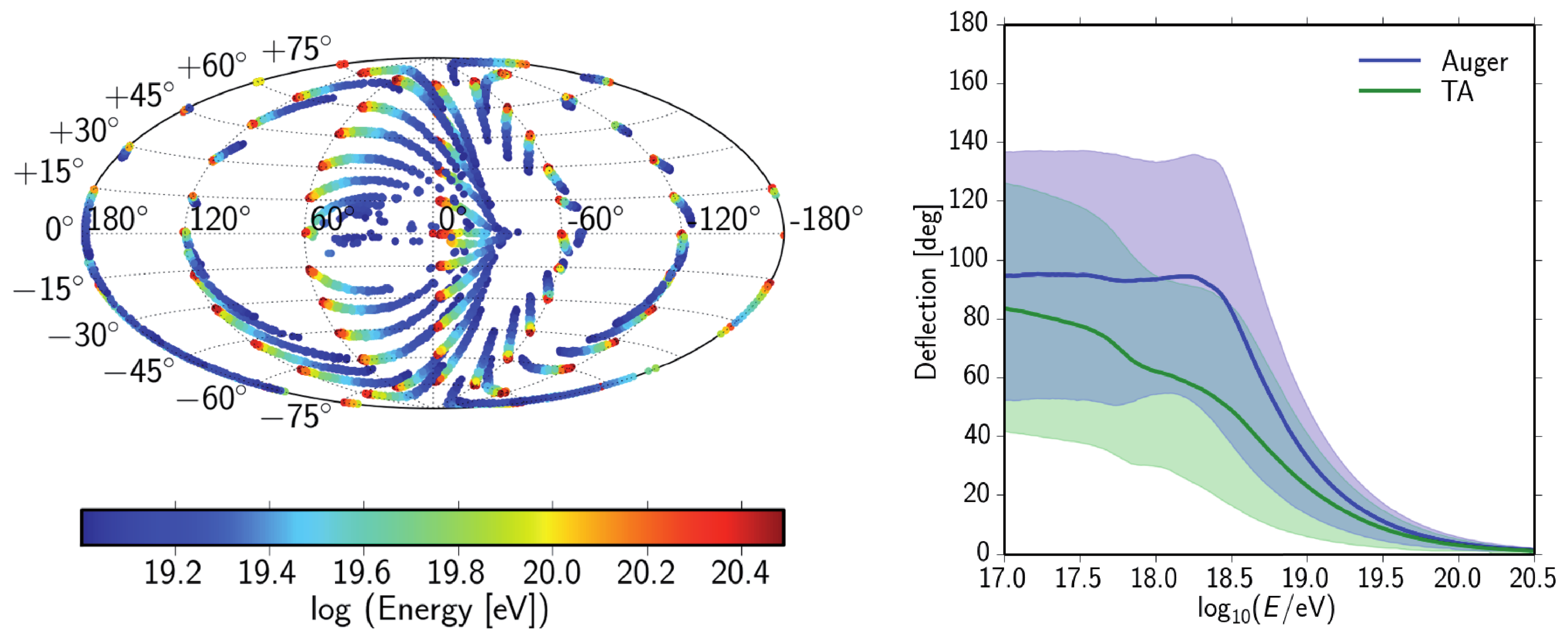}}
\caption[]{Left: Expected deflection of primary protons injected in the direction of intersecting longitude and latitude lines (dotted line) at the edge of the galaxy using the JF2012 galactic field model. The sky map is given in galactic coordinates. The color code refers to the energy of the injected proton. Right: Mean deflection of protons arriving isotropically at the edge of the galaxy using the JF2012 model. The blue line represents the mean deflection seen from the Pierre Auger Observatory site in the southern hemisphere recording particles up to 60$^\circ$ in zenith angle. The green line corresponds to the Telescope Array site in the northern hemisphere also recording particles up to 60$^\circ$ in zenith angle. The shaded area indicates the central 68\% quantile. Simulations are done using the cosmic-ray propagation code CRPropa 3.0, cf.\ Sec.\ \ref{sec:simulation}.}
\label{fig:JF12}
\end{figure}

When considering extragalactic propagation of cosmic rays, also deflections within the Milky Way may become important. Concerning \textit{galactic} magnetic fields there has been much progress in recent years. To constrain the strength of the field the best available methods are Faraday rotation measures (e.g.\ used in~\cite{Pshirkov2011}) and polarized synchrotron radiation which are both line-of-sight integrated quantities. A combination of both measurements including recent observations lead to the construction of a new galactic field model introduced in 2012 by Jansson and Farrar~\cite{Jansson:2012pc,Jansson:2012} (JF2012). One improvement compared to previous simulations is to allow for a possibility of a large-scale out-of-plane component as well as structured random fields. Typical field strengths are of the order of $\mu$G and not uniform, which implies that the angular deflection depends strongly on the observed direction as shown in Fig.\ \ref{fig:JF12}. This is important when considering anisotropies at ultra-high energy. At lower energies, e.g.\ for a primary proton of energy 1~PeV in a galactic field of 3~$\mu$G the Larmor radius is $\sim 0.4$~pc. With a diameter of the Milky Way of  $\sim 30$~kpc it is not expected to find any point sources of charged cosmic rays. At ultra-high energies there is a possibility to detect point sources and small-scale anisotropy using charged particles. A detailed knowledge of the magnetic field structure helps to interpret results from different experiments being sensitive to different parts of the sky. An example is given in Fig.\ \ref{fig:JF12} (right) selecting the Telescope Array and the Pierre Auger Observatory representing the currently largest cosmic-ray observatories for ultra-high energy particles. According to the JF2012 model, on average, the expected deflection of protons arriving isotropically at the edge of the galaxy is smaller for TA compared to the Pierre Auger Observatory. This is interesting when comparing results on anisotropy studies at both sites such as recent indications of intermediate-scale anisotropy of cosmic rays in the northern sky with TA~\cite{Abbasi:2014lda}.

\section{Simulation engines}
\label{sec:simulation}
To interpret the data collected by large-scale observatories it is necessary to develop tools that simulate the propagation of ultra-high energy cosmic rays over several orders of magnitude in energy and length scales, ranging from hundreds of Mpc down to galactic scales of order kpc including their interactions, discussed in Sects.\ \ref{sec:Prop} and \ref{sec:mag}. There has been much progress in recent years and the currently most advanced public code is CRPropa~\footnote{\url{https://crpropa.desy.de}}~\cite{Armengaud:2006fx,Kampert:2012fi}. During propagation \mbox{CRPropa} takes into account structured magnetic fields and ambient photon backgrounds including all relevant particle interactions. To enable multi-messenger analyses, secondary $\gamma$-rays and neutrinos are tracked and propagated to the observer. The code is continuously extended to handle the increasing data collected by large-scale observatories and to scan the large parameter space with high statistics, cf.\ Sec.\ \ref{sec:MPChall}. The latest version CRPropa 3.0~\cite{CRPropa3} reflects a complete redesign of the code structure, compared to the second version, to facilitate high performance computing and comprises new physical features. Simulations can be done either in a one-dimensional or three-dimensional mode. Furthermore, to take into account cosmic evolution effects in anisotropy studies and magnetic suppression in spectrum and composition studies, the latest version is augmented with a four-dimensional propagation taking into account only particles that arrive at a specific observer time. Another major improvement is the ability to take galactic deflections into account. This is realized by a lensing technique described in \cite{Bretz:2013oka} and applied in Fig.\ \ref{fig:JF12}. Photon cascades can be simulated using the electromagnetic cascade codes DINT~\cite{Lee:1996fp} or EleCa~\cite{EleCa}. Other propagation codes are e.g.\ HERMES~\cite{HERMES}, SimProp~\cite{SimProp} or TransportCR~\cite{Kalashev:2014xna}.

\section{Multiparameter challenge}
\label{sec:MPChall}

\begin{figure}
\centerline{\includegraphics[width=1.05\linewidth, draft=false]{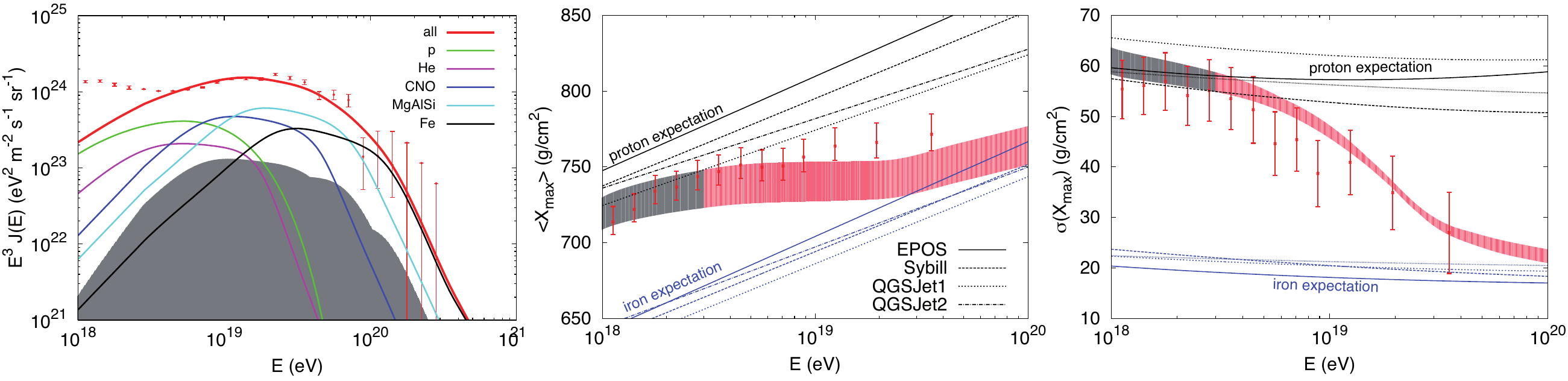}}
\caption[]{Left: Propagated cosmic-ray spectrum of protons and nuclei given an injection spectral index of $\beta = -1$ and maximum energy at the source of $E_{\rm max} = Z\times 5 \cdot 10^{18}$~eV. Solid lines indicate primary particles as labelled. The grey band shows the flux of secondaries alone. Red dots represent the measured cosmic-ray spectrum at the Pierre Auger Observatory. Center: Mean $X_{\rm max}$ as a function of energy using the same choice of parameters as for the spectrum plot. The red band illustrates the result of the simulation. The proton and iron expectations using different interaction models are indicated. The grey band denotes the energy range in which the Auger flux is not reproduced. Right: Same as the middle figure, but using its dispersion $\sigma(X_{\rm max})$. Figure adapted from \cite{Aloisio:2013hya}. References on interaction models and Pierre Auger data are given therein.}
\label{fig:DataComp}
\end{figure}

Simulation of cosmic-ray propagation involves a set of assumptions that have to be made. This stems from the fact that many unknown or uncertain parameters enter the simulation. E.g.\ sources of ultra-high energy cosmic rays are still under controversial debate, i.e.\ parameters such as total number, position, size, luminosity, composition, spectral index and emission patterns have to be estimated. Furthermore, during propagation background photon fields and magnetic field strength offer a wide parameter range. One way to disentangle information on the UHECR universe is to compare simulations with experimental data in form of suitable observables. From the observational point of view only direction and energy of the primary particle are known via the observation of extensive air showers at large-scale observatories. For example, the shape of the observed energy spectrum gives information on the sources, as well as on the propagation through the cosmic structures including the GZK effect. However, given the large parameter space in simulations, the spectrum alone can not unambiguously constrain different astrophysical scenarios and additional observables are needed. A more indirect measurement of the composition of the particle is given by the interpretation of air-shower observables such as the depth of shower maximum, usually referred to as $X_{\rm max}$ and given in g/cm$^2$, and air-shower fluctuations. Several groups have started confronting data with simulations to constrain astrophysical scenarios, e.g.\ \cite{Allard:2011aa,Taylor:2013gga,Hooper:2009fd,Taylor:2011ta,Gaisser:2013bla,Aloisio:2013hya}. These simulations indicate that typically a source with a hard spectral index is needed to explain current measurements, unless a nearby source or some additional component is assumed. An example for a fit to spectrum and composition measurements is given in Fig.\ \ref{fig:DataComp}. Given the derived hard injection spectra of $\beta = -1$ the Pierre Auger Observatory spectrum can only be fitted for energies $\gtrsim 5 \times 10^{18}$~eV. The lower energy part requires introducing a second population such as an additional class of extragalactic sources emitting mainly light elements, or a galactic cosmic-ray component~\cite{Aloisio:2013hya}. However, the latter argument requires a dominant proton fraction above $\gtrsim 10^{18}$~eV which is disfavored by upper limits on anisotropy obtained by the Pierre Auger Observatory~\cite{PAOProton} stating that the fraction of protons should not exceed $\sim 10$\% . This already indicates that it is necessary to include as much information as possible in terms of observables into the analysis.

Most commonly, comparisons to spectrum and composition data have been performed. By utilizing arrival directions as well as secondary particles (photons, neutrinos), the parameter space can be further constrained enabling a multi-messenger approach. As an example, multiplets of UHECR which exhibit energy ordering according to their angular distances relates to coherent magnetic fields. With their detection magnetic field strength can be quantified. Furthermore, high level observables such as energy-energy-correlations quantify effects of turbulent magnetic fields. So-called event-shape observables, which are being adapted from high-energy particle physics, have sensitivity to the density of sources, and probe deflections of UHECR in coherent magnetic fields. In addition, secondary messengers can be compared with observations down to the TeV energy range, refer to e.g.\ \cite{Sigl:2014jna}.

\section{Conclusion}
\label{sec:con}
The simulation of cosmic-ray propagation plays an essential role in understanding astrophysical processes at ultra-high energies. Taking into account the great wealth of data of unprecedented quality and quantity now being accumulated at large-scale observatories and sophisticated simulations based on advanced theoretical and experimental knowledge, the confrontation of data with results of simulations will lead to valuable constraints on the parameter space of theoretical models and will in this way contribute to new scientific information about the high-energy universe. It is still too early to draw decisive conclusions on astrophysical scenarios and more messengers have to be included in the analysis in a multi-messenger approach.

\section*{Acknowledgments}
It is a pleasure to thank the organizers for inviting me to the exciting 10$^{\rm th}$ Rencontres du Vietnam conference on ``Very High Energy Phenomena in the Universe'' held at the International Center of Interdisciplinary Science Education (ICISE) in the city of Quy Nhon / Vietnam. The author is grateful to stimulating discussions with David Walz who provided Figs.\ \ref{fig:density} and \ref{fig:JF12}. Financial support by the German Academic Exchange Service (DAAD) is thankfully acknowledged.

\end{document}